\documentclass[12pt]{article}
\pdfoutput=1
\usepackage[latin9]{inputenc}
\usepackage{graphicx}
\usepackage[unicode=true,
 bookmarks=false,
 breaklinks=false,pdfborder={0 0 1},backref=false,colorlinks=false]
 {hyperref}
\hypersetup{
 linktocpage}

\makeatletter

\usepackage{amssymb,amsmath,mathrsfs,epstopdf,slashed,color}
\usepackage{tikz}
\usetikzlibrary{matrix}
\usetikzlibrary{positioning}
\usepackage{fancyhdr} 
\usepackage{lastpage} 
\usepackage{extramarks} 
\usepackage{graphicx} 
\usepackage{lipsum} 
\usepackage{amsmath}
\usepackage{amsfonts}
\usepackage{amssymb}
\usepackage{latexsym}
\usepackage{color}
\usepackage{xypic}
\usepackage{cancel,slashed}
\usepackage{booktabs}
\usepackage{caption}
\usepackage{comment}
\usepackage{csquotes} 

\usepackage{graphicx}
\usepackage{placeins}
\usepackage{cite}

\allowdisplaybreaks[1]

\makeatletter

\@addtoreset{equation}{section}
\makeatother

\title{\bf Periodic Fast Radio Bursts from Axion Emission by Cosmic Superstrings}
\author{David F. Chernoff, Shing Yan Li and S.-H. Henry Tye}

\makeatother

\begin{document}
\begin{titlepage}

\setcounter{page}{0}
\begin{flushright}
\par\end{flushright}

\vskip 2.6cm 
\begin{center}
\textbf{\Large{}Periodic Fast Radio Bursts from Axion Emission by
Cosmic Superstrings}
\par\end{center}{\Large \par}

\begin{center}
\vskip 1.6cm
\par\end{center}

\begin{center}
{\large{}David F. Chernoff$^{1}$, Shing Yan Li$^{2}$ and S.-H. Henry
Tye$^{2,3,4}$}
\par\end{center}{\large \par}

\begin{center}
\vskip 0.6cm
\par\end{center}

\begin{center}
$^{1}$ Department of Astronomy, Cornell University, Ithaca, NY 14853,
USA\\
$^{2}$ Department of Physics, Hong Kong University of Science and
Technology, Hong Kong\\
$^{3}$ Jockey Club Institute for Advanced Study, Hong Kong University
of Science and Technology, Hong Kong\\
$^{4}$ Laboratory for Elementary-Particle Physics, Cornell University,
Ithaca, NY 14853, USA
\par\end{center}

\begin{center}
\vskip 0.4cm
\par\end{center}

\begin{center}
Email: \href{mailto:chernoff@astro.cornell.edu, syliah@connect.ust.hk, iastye@ust.hk}{chernoff@astro.cornell.edu, syliah@connect.ust.hk, iastye@ust.hk}
\par\end{center}

\begin{center}
\vskip 0.9cm
\par\end{center}

\begin{center}
\abstract We propose that the periodic fast radio bursts of
FRB 180916.J0158+65 are sourced by axion emission (mass $m_{a} \sim 10^{-14}$
eV) from cosmic superstrings. Some of the emitted
axions are converted to photons by magnetic fields as they travel
along the line of sight to Earth. An impulsive burst of
axion emission generates a photon signal typically lasting for
milliseconds and varying with frequency in the observed
manner. We find a range of parameters in our cosmic string network 
model consistent with the properties of FRB 180916.J0158+65. We 
suggest followup gravitational wave observations to test our model.
\par\end{center}

\begin{center}
\vspace{0.3cm}
 
\par\end{center}

\begin{flushleft}
\today 
\par\end{flushleft}

\end{titlepage}

\setcounter{page}{1} \setcounter{footnote}{0}

\tableofcontents{}

\parskip=5pt

\section{Introduction}

Fast radio bursts (FRBs) are observed as
short radio pulses typically lasting milliseconds
\cite{Lorimer:2007qn,Petroff:2019tty}. Approximately 120 have been 
detected\footnote{See catalog at \url{http://www.frbcat.org/} 
\cite{Petroff:2016tcr} and recent review \cite{Cordes:2019cmq}} and
the positions on the celestial sphere of a few have yielded
associations with extragalactic objects
\cite{Chatterjee:2017dqg,Ravi:2019alc,Bannister:2019iju}. Until this year all
FRBs were either non-repeating or repeating with unpredictable
temporal patterns \cite{Spitler:2016dmz,Amiri:2019bjk,Andersen:2019yex,Kumar:2019htf}. That changed when FRB 180916.J0158+65
(FRB 180916) was found to possess a regular period $T\simeq16.35$ days
\cite{Marcote:2020ljw,Amiri:2020gno}. This FRB is clearly associated
with a spiral galaxy at distance $r=149$ Mpc, or redshift
$z=0.03$. It emits bursts lasting milliseconds during a 4-day
window of activity followed by a 12-day quiescent period. The observed
fluence of one burst at frequency $\nu=600$ MHz for bandwidth
$\Delta\nu\sim\nu$ is $\mathcal{F}\left(\nu\right)\simeq10$ Jy-ms $\nu
\simeq6\times10^{-17}$ erg/$\mathrm{cm}^{2}$. Neutron stars, pulsars,
magnetars and white dwarfs are just some of the astronomical objects
that have been suggested as the generators of FRBs
\cite{Zhang:2020eou,Lyutikov:2020ctj,Yang:2020qxt,Levin:2020rhj,Zanazzi:2020vyp,Ioka:2020azq,Tong:2020wex,Gu:2020pyg,Mottez:2020xht}.
In this paper, we explore a different possibility involving cosmic
superstrings.

Kibble \cite{Kibble:1976sj,Kibble:1984hp} first proposed that one
dimensional topological defects might spontaneously form in grand
unified theories (GUTs) as the Universe cooled.  The defects quickly
evolve into a scaling network of horizon-size long strings and loops
of different sizes, with properties dictated largely by the string
tension $\mu$ (see \cite{Vilenkin:2000jqa} for review). The loops emit
gravitational waves and eventually
evaporate.  Strings with GUT-derived $\mu$ actively drive
cosmological perturbations whose character conflicts with the scale
invariant density perturbation spectrum favored by inflation
and observed in the cosmic
microwave background \cite{Spergel:2006hy}. Cosmic strings as fundamental
strings in string theory were first considered in the heterotic framework 
\cite{Witten:1985fp} but the tension proved
too high. More recently a consistent story of string
production after inflation was realized in the brane world scenario in
Type IIB string theory
\cite{Jones:2002cv,Sarangi:2002yt,Jones:2003da,Copeland:2003bj}.
Flux compactification and warped geometries can lower the string
tension, satisfying present day observational bounds.  Cosmic
superstrings comprise a variety of physical objects in string theory
\cite{Schwarz:1995dk}.\footnote{\label{fn2}A typical cosmic superstring network
  includes F1- and D1-strings as well as their bound states and beads
  \cite{Copeland:2003bj,Gubser:2004qj,Firouzjahi:2006vp}. The bound 
  states naturally span a range of tensions and we take $G \mu$ to be 
  a characteristic value.} 
Any discovery of them would provide solid evidence for the applicability
of string theory to nature (see
\cite{Polchinski:2004ia,Chernoff:2014cba} for review). Here, we
consider the intriguing possibility that such evidence may {\it already}
be in hand in the form of FRBs.

Cosmic superstring research has focused on strings
interacting solely by gravity but
a compelling possibility involves gravitons
and axions.\footnote{
Superstrings differ from conventional axionic strings.
Although both couple to axions, superstrings behave like local strings
while axionic strings are typically global strings having a large
spread in energy density. Our proposal requires high string frequencies
more suitable to local objects.} A string in string theory is charged
under both gravitons $h_{\mu\nu}$ and a two-form potential $A_{\mu\nu}$
with
\begin{equation}
S_{int}\propto\int d^{2}\sigma\,\sqrt{g}\left(h_{\mu\nu}g^{\alpha\beta}+\sqrt{\lambda}A_{\mu\nu}\epsilon^{\alpha\beta}\right)\partial_{\alpha}X^{\mu}\partial_{\beta}X^{\nu}\,,
\end{equation}
where $g$ is the induced metric on the worldsheet, $\epsilon^{\alpha\beta}$
is the corresponding Levi-Civita tensor and $X^{\mu}\left(\sigma\right)$
is the position of the string in spacetime. In 4-dimensional spacetime,
$A_{\mu\nu}$ is dual to an axion $a$ i.e. $\partial^{\mu}a=\epsilon^{\mu\nu\rho\sigma}\partial_{\nu}A_{\rho\sigma}$.\footnote{A realistic compactification from 10 to  4 dimensions in Type IIB string theory possesses an orientifold, in which the original two-forms $B_{\mu \nu}$ and $C_{\mu \nu}$ are projected out \cite{Copeland:2003bj,Sakellariadou:2004wq}. The axion of interest (under which a string in a warped throat is charged) is expected to be part of the complex structure (and/or K\"ahler) moduli involved in compactification. Such axions are ubiquitous in string theory. }
For massless axions, the emission of axions and gravitons
by a string have almost identical behavior, except that the amplitude
of the former is enhanced by $\sqrt{\lambda}$. In warped geometries
of flux compactification, $\lambda$ can be as large as $10^{7}$
\cite{Firouzjahi:2007dp}. Massive axions can be emitted
only when the excitation frequency $f$ exceeds the threshold energy
$\gamma_{a}m_{a}$, where $\gamma_{a}$ is a kinematic factor
and $m_{a}$ is axion mass.

The emitted axions
are converted to photons with probability $p$ when they pass through
regions of the universe with magnetic field $B$. These photons {\it are}
the FRBs we observe. To be precise, suppose the gravitational wave
flux from a cosmic superstring is $\mathcal{F}_{g}\left(f\right)$.
Then the FRB flux is given by
\begin{equation}
\mathcal{F}\left(f\right)=p\lambda\mathcal{F}_{g}\left(f\right)\Theta\left(hf-\gamma_{a}m_{a}\right)\,.
\end{equation}
The probability $p$ is negligibly small unless there is a
resonance  in which the axion mass equals the effective
photon mass derived from the plasma frequency. Therefore we require $m_{a}\simeq10^{-14}$
eV \cite{Csaki:2001jk,Mirizzi:2009iz,Mirizzi:2009nq} to match the
IGM electron density. In short, we fit the 
periodic FRB observation with a cusp from a small string loop with $G\mu \sim 10^{-8}$, $\lambda \sim 10^3$ 
and $p \sim 10^{-6}$.

In Section \ref{sec:Axion-Emission-and},
we quantify the production of FRBs from cosmic superstrings, including
the power of the axion emission and details of axion-photon conversion.
We explain the observed fluence and time dependence
of FRB 180916. In Section \ref{sec:Event-Rate-and},
we estimate the expected event rate of FRBs
by calculating the string length and string network evolution.
The rate helps delimit the possible range of model parameters
consistent with the observations. We
conclude and give remarks in Section \ref{sec:Conclusion}.

\section{Axion Emission and FRBs\label{sec:Axion-Emission-and}}

In this section, we describe how cosmic superstrings emit axions,
and how these axions are converted into FRBs.

\subsection{Power of Axion Emission}

Cosmic superstring loops emit both gravitons and axions.  The total
power of graviton emission $P_{g,tot}=\Gamma G\mu^{2}$, where $\mu$ is
the string tension and $\Gamma\simeq50$ based on numerical
calculations for representative loops. Axion emission differs from
graviton emission in two ways:
\begin{itemize}
\item The axion coupling can be much stronger than the graviton
  coupling in warped geometries. The emission power for massless
  axions is enhanced by $P_{a}\left(f\right)=\lambda
  P_{g}\left(f\right)$.
\item Axions are not massless but gain a mass $m_{a}$ from
  non-perturbative effects.  Massive axions can be emitted only when
  the emitting frequency $f$ exceeds the minimum axion energy. The
  frequency cutoff is $f\geq f_{k}=\max\left\{
  f_{1},\gamma_{a}m_{a}/h\right\} $, corresponding to the mode number
  $k=\max\left\{ 1,\gamma_{a}m_{a}l/2h\right\} $, where $l$ is the
  length of the string loop defined as $l=E/\mu$ and the fundamental
  frequency is $f_{1}=2/l$.
\end{itemize}
Cosmic superstrings are macroscopic and the production of axions requires high
frequencies characteristic of cusp and kink motions. The
gravitational emission by cusps and kinks has
been well-studied \cite{Damour:2001bk,Damour:2004kw}. The gravitational wave
amplitude $h_{m}$ with mode number $m$ in the direction of the cusp
is asymptotically $h_{m}\propto m^{-4/3}$.\footnote{Due to the nonlinear periodic motion the loop can emit gravitons at
all modes.} The power per solid angle
is $dP_{m}/d\Omega\propto\left(f_{m}h_{m}\right)^{2}\propto m^{-2/3}$
and $\sum_{m}dP_{m}/d\Omega$
diverges in the exact direction of the cusp.
In fact, the emitted gravitons of the $m$-th mode form a narrow
beam with width $\Theta_{m}=m^{-1/3}$ and solid angle $\Omega_{m}\simeq\pi\Theta^{2}$.
The power within $\Omega_{m}$ is $P_{m}\simeq\left(dP_{m}/d\Omega\right)\Omega_{m}\propto m^{-4/3}$.
Assuming the asymptotic limit for all $m$
gives $P_{g,tot}=\sum_{m}P_{m}=P_{1}\zeta\left(4/3\right)$
and $P_{m}=\Gamma G\mu^{2}m^{-4/3}/\zeta\left(4/3\right)$. The same
procedure for kinks leads to $P_{m}\propto m^{-5/3}$, implying
that the power at large $m$ from kinks is subdominant when cusps are present.
Therefore, below we focus
on gravitons and axions emitted by cusps.

The FRB appears smooth at frequency $\nu$ over a bandwidth $\Delta\nu\sim\nu$.
The gravitational wave power
within a frequency window $\left[f_{m},f_{2m}\right]$ is
$P_{m,2m}=\sum_{i=m}^{2m}P_{i}\simeq3m^{-1/3}\left(1-2^{-1/3}\right)P_{1}$
and the average brightness
is $dP_{m,2m}/d\Omega_{2m} \equiv P_{m,2m}/\Omega_{2m}=\Gamma G\mu^{2}\frac{3\cdot2^{2/3}\left(1-2^{-1/3}\right)}{\zeta\left(4/3\right)\pi}m^{1/3}$.
The average flux received at Earth is $F_{m,2m}=\left(1/r^{2}\right)dP_{m,2m}/d\Omega_{2m}$
and the gravitational wave fluence is $\mathcal{G}\left(\nu\right)=F_{m,2m}\Delta t$,
where $\Delta t=1/\nu$ is the characteristic time scale for the cusp
at frequency $\nu$. Finally the axion fluence is $\lambda\mathcal{G}\left(\nu\right)$
and the FRB fluence is $\mathcal{F}\left(\nu\right)=p\lambda\mathcal{G}\left(\nu\right)$
if $\nu>\gamma_{a}m_{a}/h$.

Let us apply this result to FRB 180916. If the source is a
cosmic superstring loop with non-relativistic center of mass motion, its
period $T=16.35$ days corresponds to the string length
$l=2T=8.5\times10^{16}$ cm.  Assume one cusp per period points towards
us i.e. $n_{c}=1$.\footnote{The simplest loop forms two cusps per period but these generally
point in different directions.} The FRB frequency $\nu=600$ MHz
corresponds to the mode number $m=\nu T=8.5\times10^{14}$, which is
much larger than the mode number cutoff
$k\left(l\right)=3.4\times10^{6}$ for $\gamma_{a} m_{a}=10^{-14}$ eV.
The cosmological redshift is negligible. The fluence is
\begin{equation}
\mathcal{F}\left(\nu\right)=p\lambda\Gamma G\mu^{2}\frac{3\cdot2^{2/3}\left(1-2^{-1/3}\right)}{\pi\zeta\left(4/3\right)}\frac{m^{1/3}}{r^{2}\nu}=1200p\lambda\left(G\mu\right)^{2}\,\mathrm{erg/cm^{2}}\,.
\end{equation}
The observed fluence
$\mathcal{F}\left(\nu\right)=6\times10^{-17}\:\mathrm{erg/cm^{2}}$
implies the constraint
$p\lambda\left(G\mu\right)^{2}=5\times10^{-20}$ on
the important string theory parameters $\mu$ and $\lambda$, and
the axion-photon conversion probability $p$.

If the center of mass of the
cosmic superstring loop moves with Lorentz factor
$\gamma_{s}\gg1$ towards the Earth, all frequencies are Doppler 
shifted such that in the center-of-mass frame, the string length is 
$l'=\left(2\gamma_{s}\right)2T$ and the emitting frequency is 
$\nu'=\nu/2\gamma_{s}$. This leads to the same $m$. Moreover, the fluence is enhanced
by relativistic beaming with factor of $8\gamma_{s}^{3}$. The
fluence constraint becomes $\gamma_{s}^{4}p\lambda\left(G\mu\right)^{2}=3\times10^{-21}$.

\subsection{Axion-Photon Conversion}

When axions pass through a magnetic field $B$ they may be converted
into photons. Such a process requires an axion-photon coupling,
which can be described by the following effective theory
\begin{equation}
\mathcal{L}=-\frac{1}{4}F^{\mu\nu}F_{\mu\nu}-\frac{1}{4f_{a}}aF^{\mu\nu}\bar{F}_{\mu\nu}-\frac{1}{2}\partial_{\mu}a\partial^{\mu}a-\frac{1}{2}m_{a}^{2}a^{2}\,,
\end{equation}
where $\bar{F}_{\mu\nu}$ is the dual of the electromagnetic field strength $F_{\mu\nu}$ and $f_{a}$ is
the effective axion decay constant. When transverse $B$ is present, axion-photon
oscillations happen with angular frequency $\omega_{osc}\sim B/f_{a}$;
in dimensional units with $\chi = (B/{\rm nG})(10^{10}\,{\rm GeV}/f_{a})$
  we have $\omega_{osc}= 1.1 \times 10^{-14} \chi $ rad/s,
period $T_{osc} = 6.0 \times 10^{14}/\chi$ s,
wave number $k_{osc} = 1.1 \chi $ Mpc$^{-1}$
and wave length $\lambda_{osc} = 5.8/\chi$ Mpc
\cite{Raffelt:1987im}.

This process has been examined in the homogeneous early universe
\cite{Yanagida:1987nf,Mirizzi:2009iz,Mirizzi:2009nq}.  The variation
of the electron density with epoch
is equivalent to a time-dependent effective photon
mass $m_{\gamma}\left(t\right)$.  If the axion is emitted out of
resonance and passes through resonance as the universe expands then some
axions are converted to photons. The total distance traveled needs to
be much larger than the oscillation wavelength. If we apply the same
logic to FRB 180916 resonance must occur at
$z\lesssim0.03$. Numerically, $m_{a}\simeq
m_{\gamma}\left(t_{0}\right)\simeq10^{-14}$ eV, where $t_{0}$ is the
age of universe.  The photon mass is $m_{\gamma}\left(t\right)\simeq
m_{a}\left(1+3H\left(t_{0}-t\right)/2\right)$ for $t$ near $t_{0}$, where $H$ is the Hubble
constant. The Landau-Zener solution 
\cite{Mirizzi:2009iz,Mirizzi:2009nq,Moroi:2018vci} for passing through
resonance at constant rate gives the
probability $p$ for conversion of axion to photon:
\begin{equation}
p\simeq \frac{\pi\left(B/f_{a}\right)^{2}\nu}{9m_{a}^{2}H}\,.
\end{equation}
Using typical values $\chi = 1$ (e.g. $B=1$ nG and $f_{a}=10^{10}$
GeV) we estimate $p\simeq4.5\times10^{-5}$. The actual situation is,
of course, more complicated. The transverse field component
and the electron density may be inhomogeneous along the line of sight,
impacting the resonance condition and the characteristic oscillation
length scale $\lambda_{osc}$. We expect plasma outflows from AGN and star
forming galaxies to carry magnetic fields into the IGM.
In general, $\lambda_{osc}$ decreases and $p$ increases as
$B$ increases so axion-photon conversion could be sensitive
to the small scale inhomogeneous injection of fields.
For now we regard $p$ as a parameter that
could span a wide range of values much less than 1.

The burst of axions has intrinsic width $1/\nu$ at frequency of
observation $\nu$, nearly a delta function in time.  Despite the
uncertainty in the size of $p$, the conversion process imprints a minimum width to
the photon burst.  For FRB 180916, the frequency $\nu=600$ MHz
corresponds to axions moving at $\gamma=h\nu/m_{a}\gg1$, or
$1-\beta\simeq1/(2 \gamma^2) = 8\times10^{-18} (600\, {\rm MHz}/\nu)^2
(m_{a}/10^{-14}\, \rm{eV})^2$.  If some axions are converted to photons
later than others by time $\delta t$ then their arrival times
will lag by $\left(1-\beta\right)\delta t$. The axion-photon
oscillation implies a minimum conversion time scale $\delta t=
T_{osc}$ and a characteristic observed burst duration
$t_{burst}=\left(1-\beta\right)\delta t=4 (600\, {\rm MHz}/\nu)^2
(m_{a}/10^{-14}\, \rm{eV})^2/\chi$ ms. The minimum temporal spread
for the FRB is milliseconds. The actual spread will be supplemented
by dispersion (DM) and scattering (SM) effects of the medium
\cite{Xu:2016vjj}.

The DM time delay is a strict function of frequency that can
be exactly removed for a pulsed, broadband radio source by fitting the
known form of the delay $\propto 1/\nu^2$. In practice, the
observer minimizes the variance of the observed and theoretically
delayed signal. In the axion-photon case
the size of the characteristic delay varies $\propto 1/\nu^2$. Part
will be absorbed when the delay of the FRB is
fitted. For FRB 180916 $DM=349.02$
pc-cm$^{-3}$ implies $0.4$ s delays at 600 MHz.  The width of the
sub-bursts after fitting is about $5$ ms. Fitting cannot shrink the
width of the burst back to a delta function which is governed by
the quantum mechanical transition from axion to photon.

There are aspects of the signals that are not well-understood. In particular, 
the above simple model does not explain the 4-day active period. 
The loop might have multiple kinks and cusps. 
Although a kink signal is sub-dominant, it is more 
likely to be observed because the emission is fan-like instead of point-like as is the case 
of a cusp emission.
Gravitational lensing might magnify the signal and generate multiple bursts 
arriving at different times.

\section{Event Rate and Estimation of Parameters\label{sec:Event-Rate-and}}

From above, we see that a large range of parameters can satisfy the
fluence constraint for FRB 180916. We now impose the condition that
the event rate for FRBs like 180916 should not be unobservably low.

\subsection{String Length Evolution}

\noindent A single
cosmic superstring loop emits axions at mode numbers $m\geq k\left(l\right)$
only. When emission per mode scales like that of a cusp the loop
has total axion power $P_{a,tot}=\lambda P_{1}\sum_{n=k}^{\infty}n^{-4/3}\simeq\lambda k^{-1/3}P_{g,tot}$.
As shown above, the observed emission from FRB 180916 requires $k\gg1$.
We focus on the case that the
axion emission is large compared to gravitational emission:
$\lambda k^{-1/3}>1$. In this limit axion emission,
not graviton emission, dominates the evolution in length of the string.
Writing $k=\tilde{k}l$, where $\tilde{k}=\gamma_{a}m_{a}/2h$, 
the length evolves according to
\begin{equation}
\frac{dl}{dt}=-\Gamma G\mu\frac{\lambda}{\left(\tilde{k}l\right)^{1/3}}\Rightarrow t-t_{b}=\frac{3\tilde{k}^{1/3}}{4\Gamma G\mu\lambda}\left(l_{b}^{4/3}-l^{4/3}\right)\,,\label{eq:evolution}
\end{equation}
where $t_{b}$ is the time when the loop was born and $l_{b}$
is the length at that time.

The cosmic superstring network follows the scaling attractor solution
in deep radiation or matter era \cite{Kibble:1976sj,Kibble:1984hp}.\footnote{The evaporation of loops by the combined action of axions and gravitons
  does not alter the Velocity One Scale model solution.}
For simplicity, we assume scaling applies at all times in the cosmological model
so that the loop size is always proportional to horizon size
$l_{b}=\alpha t_{b}$. The condition that a loop born at $t_{b}$ hasn't
yet evaporated by the current epoch is $t_{b}\geq t_{b,min}$.
There are large and small $\alpha$ limiting cases:
if $\alpha\gg(\Gamma G\mu\lambda)^{3/4}/\gamma_{a}^{1/4}$ then
\begin{equation}
  t_{b,min}=\frac{1}{\alpha}\left(\frac{2h}{\gamma_{a}m_{a}}\right)^{1/4}
  \left(\Gamma G\mu\lambda t_{0}\right)^{3/4}\,.\label{eq:mintb}
\end{equation}
implying $t_{b,min} \ll t_{0}$
and if $\alpha\ll\Gamma G\mu\lambda$ then $t_{b,min}\approx t_{0}$.

In the $\Lambda$CDM model, one of the crucial factors affecting the
string network evolution is whether the universe is in radiation or
matter era. The time of equipartition $t_{eq}=1.5\times10^{12}$
s. If $t_{b,min}>t_{eq}$, the string loops we observe are all produced
in matter-dominated era. If $t_{b,min}<t_{eq}$, some small string
loops we observe are produced in radiation-dominated era. For large
$\alpha$, the latter condition is equivalent to an upper bound of
string tension $\mu$:
\begin{equation}
\alpha^{-4/3}\gamma_{a}^{-1/3}G\mu\lambda<8.4\times10^{-4}\,.
\end{equation}
where we have taken $m_{a}=10^{-14}$ eV.
All these results will be used in deriving the string network
evolution.

\subsection{String Size Distribution}

To infer the event rate, we estimate the number densities $n$ of
loops of different sizes with the Velocity One Scale (VOS) model \cite{Martins:1995tg,Battye:1997hu,Pogosian:1999np,Martins:2000cs}
and loop production characterized in complementary analytic
\cite{Polchinski:2007rg,Polchinski:2007qc} and
numerical treatments
\cite{Ringeval:2005kr,Lorenz:2010sm,BlancoPillado:2011dq,Blanco-Pillado:2013qja}.\footnote{There remain important differences in the number of small loops inferred by simulation. Our approach follows
  \cite{Chernoff:2009tp,Chernoff:2017fll}.}
About $f=80\%$ of
the energy goes to small loops with $\alpha\simeq20\left(G\mu\right)^{1.2}$
in radiation era, and $\alpha\simeq20\left(G\mu\right)^{1.5}$ in
matter era \cite{Polchinski:2007rg,Polchinski:2007qc}. In both cases
$\alpha\ll\Gamma G\mu\lambda$. The remaining $f=20\%$
goes to large loops with $\alpha\simeq0.1\gg\Gamma G\mu\lambda$ as generally seen in simulations. The birth rate density of string loops
at time $t_{b}$ during scaling is
\begin{equation}
\frac{dn}{dt_{b}}\left(t_{b}\right)=\frac{fA}{\alpha t_{b}^{4}}\,,
\end{equation}
where $A=7.65$ in radiation era and $A=0.55$ in matter
era \cite{Chernoff:2014cba,Chernoff:2017fll}. When the universe expands,
the loop number density $\propto a^{-3}$ for scale factor $a\left(t\right)$.
Performing a change of variable the density of loops
with size $l$ at time $t$,
\begin{equation}
l\frac{dn}{dl}\left(l,t\right)=\frac{fA}{\alpha t_{b}^{4}}\frac{a^{3}\left(t_{b}\right)}{a^{3}\left(t\right)}\frac{l}{\left|dl/dt_{b}\right|}\,,
\end{equation}
where $t_{b}$ and $\left|dl/dt_{b}\right|$ are determined
from Eq. (\ref{eq:evolution}). To be precise,
\begin{equation}
\left|\frac{dl}{dt_{b}}\right|=\left(\frac{\alpha}{\lambda/(\tilde{k}l_{b})^{1/3}}+\Gamma G\mu\right)\frac{\lambda}{\left(\tilde{k}l\right)^{1/3}}\,.
\end{equation}
For cosmic superstrings, the Universe's mean loop density is further enhanced
by a factor $\mathcal{G}\sim10^{3}$, which is due to a combination
of multiple throats in flux compactification, multiple string species
and low intercommutation probability of superstrings \cite{Chernoff:2014cba,Chernoff:2017fll}.
The superstring size distribution is $\mathcal{G}l\left(dn/dl\right)\left(l,t\right)$.

\subsection{Estimation of Parameters}

If FRB 180916 comes from a cosmic superstring, the loop density
at the corresponding length and time should not be too low else
detection will be highly improbable.  This condition
constrains the allowed range of parameters tightly.

How can a lower bound of the loop density be set? The
current FRB detectors can measure FRBs with fluence as small as $0.1\mathcal{F}\left(\nu\right)$. Sources within distance
$r_{max}=10^{1/2}r$ could have been seen by existing surveys.
Write $V_{max}=(4/3)\pi r_{max}^3$.
The expected number of superstrings
$N\left(l,\nu\right)$ with length $l$ that can
emit detectable FRBs with frequency $\nu$ in an all sky survey is
\begin{equation}
N\left(l,\nu\right)\simeq\mathcal{G}l\frac{dn}{dl}V_{max}\frac{\Omega}{4\pi}\,,
\end{equation}
where the solid angle of the cusp beam is $\Omega=\pi/m^{2/3}$.
As a conservative choice, we impose $N>10^{-5}$.
For work below we assume $\gamma_{a}=1$.

In conventional cosmic string scenarios, large loops source
the potentially observable
signals. Axions increase the
rate of evaporation of all loops compared to gravitons alone.
Consequently, suppression of $dn/dl$ occurs
for both small and large loops and the
suppression of signals is greater for the large loops.
Hence, we must consider both types as potential FRB sources.
In general, small loops are born with highly relativistic motions
$\gamma_{s}\gg1$. In such cases, the string length\footnote{The 
  string length in the lab frame $l$ is related to
  that in the rest frame $l'$ by $l=\gamma_{s}l'$, i.e. the so-called 
  loop invariant length transforms like energy.} 
is $l=\left(2\gamma_{s}^{2}\right)2T$ and $\Omega$ is suppressed 
by relativistic beaming with factor of $1/2\gamma_{s}^2$. Overall, 
when comparing to the non-relativistic case, $N$ 
is enhanced by factor of $\left(2\gamma_{s}^2\right)^{1/3}$,
giving modestly larger allowed parameter ranges.
We will assume $\gamma_{s}=1$, thus $l=8.5\times10^{16}$ cm, 
$m=8.5\times10^{14}$ and $k\left(l\right)=3.4\times10^{6}$ .

\begin{figure}
\begin{centering}
\includegraphics[width=0.5\paperwidth]{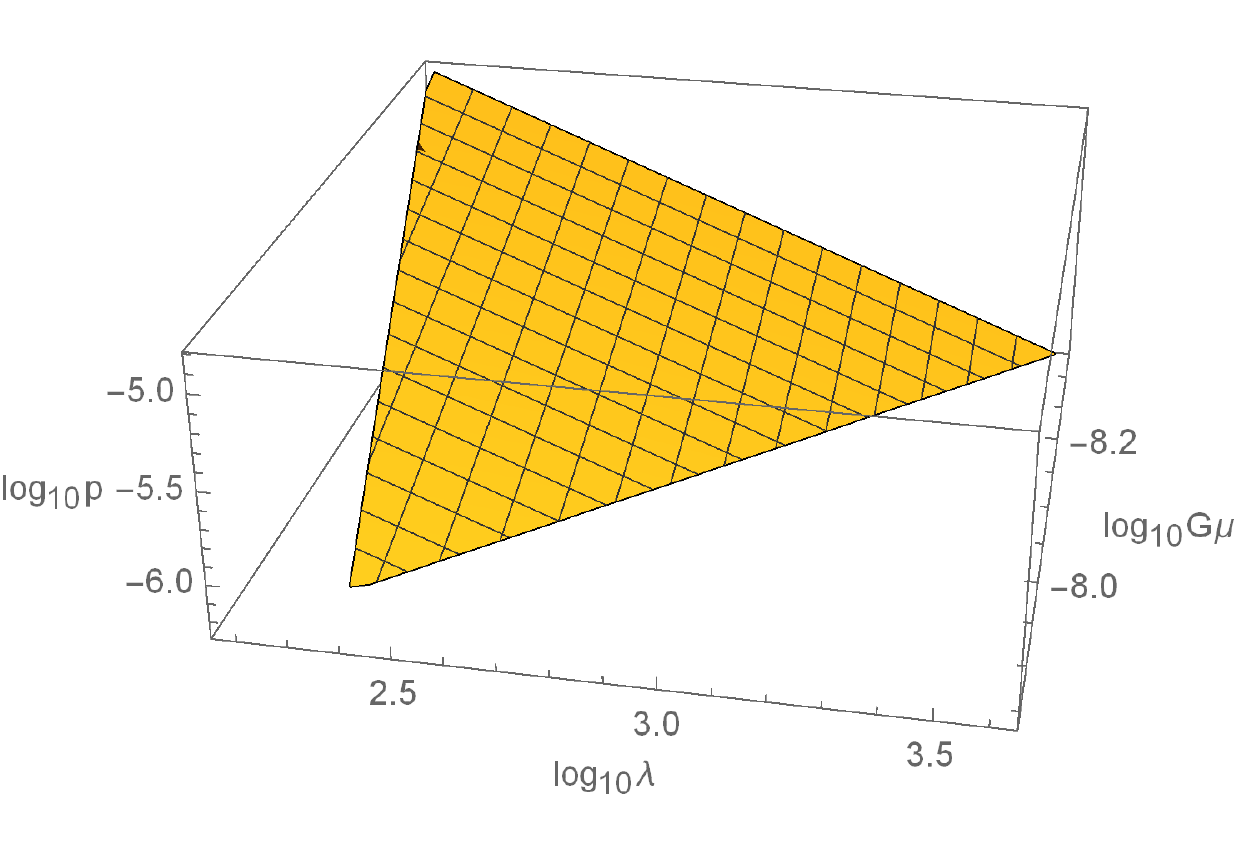}
\par\end{centering}
\caption{Range of allowed parameters $p,\lambda,G\mu$ if FRB 180916 is from
a recently born small superstring loop with $\alpha\simeq20\left(G\mu\right)^{1.5}$.
The parameters are allowed if they give the observed fluence $\mathcal{F}\left(\nu\right)=6\times10^{-17}$
erg/$\mathrm{cm}^{2}$ and expected number of such loop $N>10^{-5}$
within radius $10^{1/2}r=470$ Mpc, and axion emission dominates the
string evolution. For simplicity we assume a non-relativistic loop,
and relativistic loops give larger range of parameters.\label{fig:Range-of-allowed}}

\end{figure}

First consider the possibility of a small loop. If the string is a
recently born small loop, we have $\alpha=20\left(G\mu\right)^{1.5}$
and $t_{b}\simeq t_{0}$. To be consistent, we must require $l_{b}\left(t_{0}\right)>l$,
which implies a minimum tension $G\mu>4.7\times10^{-9}$.\footnote{
  This exceeds the constraint from pulsar timing array $G\mu<10^{-9}$ for
  strings coupled only to gravity\cite{Sousa:2016ggw}. With $\lambda \gg 1$
  the loop density and stochastic gravitational wave background are lowered,
  and the bounds from null results of gravitational wave measurements
  are much relaxed. Moreover, such
  observations bound the characteristic 
  tension only (see Footnote \ref{fn2}). Here $G\mu$ can be higher
  if the loop is a bound state.} 
We have $\lambda>k\left(\alpha t_{0}\right)^{1/3}=2\times10^{6}\left(G\mu\right)^{0.5}\gtrsim150$
if axion emission dominates. Assuming so, we can approximate $\left|dl/dt_{b}\right|=\Gamma G\mu\lambda/k\left(l\right)^{1/3}$.
The condition on $N$ then implies $\left(G\mu\right)^{2.5}\lambda<6.3\times10^{-18}$,
hence $G\mu<1.4\times10^{-8}$. The minimum tension yields $\lambda<4.2\times10^{3}$.
The constraint from the observed fluence implies $5.4\times10^{-7}<p<1.5\times10^{-5}$,
which is close to our estimation in Section \ref{sec:Axion-Emission-and}.
Overall, the allowed range of parameters is plotted in Figure \ref{fig:Range-of-allowed}.
The range is narrow but possible.

If the string was a large loop in the past, we have $\alpha=0.1$.
Since the string loop size is much smaller than the horizon size nowadays,
it must be born in the early universe with length $l_{b}=\alpha t_{b}$.
We estimate that it was born at $t_{b,min}$, which is given by Eq.
(\ref{eq:mintb}). We further assume axion emission dominates and
$t_{b,min}$ is in radiation era, which means $G\mu\lambda<4\times10^{-5}$.
We approximate the scale factor as $a\left(t_{b}\right)/a\left(t_{0}\right)=\left(t_{b}/t_{eq}\right)^{1/2}\left(t_{eq}/t_{0}\right)^{2/3}$
and $\left|dl/dt_{b}\right|=\alpha\left(l_{b}/l\right)^{1/3}$. The
constraint on $N$ then gives $G\mu\lambda<3.0\times10^{-7}$. The
lower limit of $\lambda$ is given by the axion emission domination
$\lambda>k\left(l_{b}\right)^{1/3}$, which implies $\lambda>3\times10^{6}\left(G\mu\right)^{1/3}$
and $G\mu<1.8\times10^{-10}$. Combining with the observed fluence,
we get $p>10^{-3}$. If we assume that $t_{b,min}$
is in matter era, we require $G\mu\lambda>4\times10^{-5}$ which will
contradict with the constraint on $N$.

Since the allowed $p$ is exceptionally high in the large loop case,
the observed FRB probably comes from a recently born small
loop. The estimated typical parameters are $G\mu\sim10^{-8}$,\footnote{It 
falls within the range of $(p,q)$ string tension \cite{Copeland:2003bj,Firouzjahi:2005dh} in the 
KKLMMT scenario \cite{Kachru:2003sx}.} $\lambda\sim10^{3}$
and $p\sim10^{-6}$ for $\gamma_{s}=1$ and $N>10^{-5}$. There are
considerable observational and/or theoretical uncertainties in ${\cal F}(\nu), n_{c}, 
\Gamma$, $m_{a}$, ${\cal G}$ and number density of small loops.

\section{Conclusion\label{sec:Conclusion}}

We have demonstrated that observable periodic FRBs such as FRB 180916
can be produced from axion emission by low-tension cosmic superstrings,
following by axion-photon conversion. We have developed a simple model
of cosmic superstrings coupled to both gravitons and axions
with large axion coupling. The evolution
of strings and the nature of the emission differs from previously
studied models. We
have found a range of parameters that gives the observed fluence $\mathcal{F}\left(\nu\right)$,
and estimated a small but potentially detectable number of sources like
FRB 180916. Of the options considered
we favor that the FRB is produced by a recently
born small loop with $\alpha\simeq20\left(G\mu\right)^{1.5}$
as opposed to an older loop born with large size. The
proposed tension $G\mu\sim10^{-8}$ is quite high but self-consistent
since the upper bounds from experiments probing gravitational
waves are relaxed for large $\lambda$. The process of axion-photon conversion
provides a not fully quantitative explanation of the time dependence
of FRBs, specifically the intrinsic burst duration.
All the mechanisms outlined here will function for large
string loops. The FRB repetition time is related to the
fundamental period of the loop which may exceed the
time span of observations to date.

The results here can be improved with more data and theoretical
studies. Cosmic superstrings may generate observable gravitational wave bursts.
Writing $G\mu=10^{-8}\mu_{-8}$, in our model the
gravitational wave burst fluence at frequency $\nu_{1}$ Hz associated
with FRB 180916 is
\begin{equation}
\mathcal{G}\left(\nu_{1}\,\mathrm{Hz}\right)=8.5\times10^{-8}\mu_{-8}^{2}\nu_{1}^{-2/3}\,\mathrm{erg/cm^{2}}\,.
\end{equation}
We suggest searching for gravitational wave emission
at the source position of the FRB. The signal-to-noise ratio can be enhanced
by folding the signal at the known periodicity.

If FRBs are confirmed experimentally to be generated by cosmic
superstrings they will provide important insights into string theory
and the cosmological history of the Universe.

\section*{Acknowledgment}

We thank Ira Wasserman, George Smoot and Tom Broadhurst for discussions. DFC acknowledges 
that this material is based upon work supported by
the National Science Foundation under Grant No. 1417132. SHHT is supported
by the AOE grant AoE/P-404/18-6 issued by the
Research Grants Council (RGC) of the Government of the Hong Kong SAR.

\bibliographystyle{utphys}
\bibliography{refs}

\providecommand{\href}[2]{#2}\begingroup\raggedright\begin{thebibliography}{10}

\bibitem{Lorimer:2007qn}
D.~R. Lorimer, M.~Bailes, M.~A. McLaughlin, D.~J. Narkevic and F.~Crawford,
  {\em {A bright millisecond radio burst of extragalactic origin}}, Science
  {\bf 318} (2007) 777
[\href{http://www.arXiv.org/abs/0709.4301}{{\tt 0709.4301}}].

\bibitem{Petroff:2019tty}
E.~Petroff, J.~W.~T. Hessels and D.~R. Lorimer,  {\em {Fast Radio Bursts}},
  Astron. Astrophys. Rev. {\bf 27} (2019), no.~1, 4
[\href{http://www.arXiv.org/abs/1904.07947}{{\tt 1904.07947}}].

\bibitem{Petroff:2016tcr}
E.~Petroff, E.~D. Barr, A.~Jameson, E.~F. Keane, M.~Bailes, M.~Kramer,
  V.~Morello, D.~Tabbara and W.~van Straten,  {\em {FRBCAT: The Fast Radio
  Burst Catalogue}}, Publ. Astron. Soc. Austral. {\bf 33} (2016) e045
[\href{http://www.arXiv.org/abs/1601.03547}{{\tt 1601.03547}}].

\bibitem{Cordes:2019cmq}
J.~M. Cordes and S.~Chatterjee,  {\em {Fast Radio Bursts: An Extragalactic
  Enigma}}, Ann. Rev. Astron. Astrophys. {\bf 57} (2019) 417--465
[\href{http://www.arXiv.org/abs/1906.05878}{{\tt 1906.05878}}].

\bibitem{Chatterjee:2017dqg}
S.~Chatterjee {\em et al.},  {\em {The direct localization of a fast radio
  burst and its host}}, Nature {\bf 541} (2017) 58
[\href{http://www.arXiv.org/abs/1701.01098}{{\tt 1701.01098}}].

\bibitem{Ravi:2019alc}
V.~Ravi {\em et al.},  {\em {A fast radio burst localized to a massive
  galaxy}}, Nature {\bf 572} (2019), no.~7769, 352--354
[\href{http://www.arXiv.org/abs/1907.01542}{{\tt 1907.01542}}].

\bibitem{Bannister:2019iju}
K.~W. Bannister {\em et al.},  {\em {A single fast radio burst localized to a
  massive galaxy at cosmological distance}},
\href{http://www.arXiv.org/abs/1906.11476}{{\tt 1906.11476}}.

\bibitem{Spitler:2016dmz}
L.~G. Spitler {\em et al.},  {\em {A Repeating Fast Radio Burst}}, Nature {\bf
  531} (2016) 202
[\href{http://www.arXiv.org/abs/1603.00581}{{\tt 1603.00581}}].

\bibitem{Amiri:2019bjk}
{CHIME/FRB} Collaboration, M.~Amiri {\em et al.},  {\em {A Second Source of
  Repeating Fast Radio Bursts}}, Nature {\bf 566} (2019), no.~7743, 235--238
[\href{http://www.arXiv.org/abs/1901.04525}{{\tt 1901.04525}}].

\bibitem{Andersen:2019yex}
{CHIME/FRB} Collaboration, B.~C. Andersen {\em et al.},  {\em {CHIME/FRB
  Discovery of Eight New Repeating Fast Radio Burst Sources}}, Astrophys. J.
  {\bf 885} (2019), no.~1, L24
[\href{http://www.arXiv.org/abs/1908.03507}{{\tt 1908.03507}}].

\bibitem{Kumar:2019htf}
P.~Kumar {\em et al.},  {\em {Faint Repetitions from a Bright Fast Radio Burst
  Source}}, Astrophys. J. {\bf 887} (2019), no.~2, L30
[\href{http://www.arXiv.org/abs/1908.10026}{{\tt 1908.10026}}].

\bibitem{Marcote:2020ljw}
B.~Marcote {\em et al.},  {\em {A repeating fast radio burst source localised
  to a nearby spiral galaxy}}, Nature {\bf 577} (2020), no.~7789, 190--194
[\href{http://www.arXiv.org/abs/2001.02222}{{\tt 2001.02222}}].

\bibitem{Amiri:2020gno}
{CHIME/FRB} Collaboration, Amiri {\em et al.},  {\em {Periodic activity from a
  fast radio burst source}},
\href{http://www.arXiv.org/abs/2001.10275}{{\tt 2001.10275}}.

\bibitem{Zhang:2020eou}
B.~Zhang,  {\em {Fast Radio Bursts from Interacting Binary Neutron Star
  Systems}}, Astrophys. J. {\bf 890} (2020), no.~2, L24
  [\href{http://www.arXiv.org/abs/2002.00335}{{\tt 2002.00335}}],
[Astrophys. J. Lett.890,L24(2020)].

\bibitem{Lyutikov:2020ctj}
M.~Lyutikov, M.~Barkov and D.~Giannios,  {\em {FRB-periodicity: mild pulsar in
  tight O/B-star binary}},
\href{http://www.arXiv.org/abs/2002.01920}{{\tt 2002.01920}}.

\bibitem{Yang:2020qxt}
H.~Yang and Y.-C. Zou,  {\em {Orbital-induced spin precession as an origin of
  periodicity in periodically-repeating fast radio bursts}},
\href{http://www.arXiv.org/abs/2002.02553}{{\tt 2002.02553}}.

\bibitem{Levin:2020rhj}
Y.~Levin, A.~M. Beloborodov and A.~Bransgrove,  {\em {Precessing flaring
  magnetar as a source of repeating FRB 180916.J0158+65}},
\href{http://www.arXiv.org/abs/2002.04595}{{\tt 2002.04595}}.

\bibitem{Zanazzi:2020vyp}
J.~J. Zanazzi and D.~Lai,  {\em {Periodic Fast Radio Bursts with Neutron Star
  Free/Radiative Precession}},
\href{http://www.arXiv.org/abs/2002.05752}{{\tt 2002.05752}}.

\bibitem{Ioka:2020azq}
K.~Ioka and B.~Zhang,  {\em {A Binary Comb Model for Periodic Fast Radio
  Bursts}},
\href{http://www.arXiv.org/abs/2002.08297}{{\tt 2002.08297}}.

\bibitem{Tong:2020wex}
H.~Tong, W.~Wang and H.~G. Wang,  {\em {Periodicity in fast radio bursts due to
  forced precession by a fallback disk}},
\href{http://www.arXiv.org/abs/2002.10265}{{\tt 2002.10265}}.

\bibitem{Gu:2020pyg}
W.-M. Gu, T.~Yi and T.~Liu,  {\em {A Neutron Star-White Dwarf Binary Model for
  Periodic Fast Radio Bursts}},
\href{http://www.arXiv.org/abs/2002.10478}{{\tt 2002.10478}}.

\bibitem{Mottez:2020xht}
F.~Mottez, G.~Voisin and P.~Zarka,  {\em {Repeating fast radio bursts caused by
  small bodies orbiting a pulsar or a magnetar}},
\href{http://www.arXiv.org/abs/2002.12834}{{\tt 2002.12834}}.

\bibitem{Kibble:1976sj}
T.~W.~B. Kibble,  {\em {Topology of Cosmic Domains and Strings}}, J. Phys. {\bf
  A9} (1976)
1387--1398.

\bibitem{Kibble:1984hp}
T.~W.~B. Kibble,  {\em {Evolution of a system of cosmic strings}}, Nucl. Phys.
  {\bf B252} (1985) 227
[Erratum: Nucl. Phys.B261,750(1985)].

\bibitem{Vilenkin:2000jqa}
A.~Vilenkin and E.~P.~S. Shellard, {\em {Cosmic Strings and Other Topological
  Defects}}.
\newblock Cambridge University Press,
2000.
\newblock

\bibitem{Spergel:2006hy}
{WMAP} Collaboration, D.~N. Spergel {\em et al.},  {\em {Wilkinson Microwave
  Anisotropy Probe (WMAP) three year results: implications for cosmology}},
  Astrophys. J. Suppl. {\bf 170} (2007) 377
[\href{http://www.arXiv.org/abs/astro-ph/0603449}{{\tt astro-ph/0603449}}].

\bibitem{Witten:1985fp}
E.~Witten,  {\em {Cosmic Superstrings}}, Phys. Lett. {\bf 153B} (1985)
243--246.

\bibitem{Jones:2002cv}
N.~T. Jones, H.~Stoica and S.~H.~H. Tye,  {\em {Brane interaction as the origin
  of inflation}}, JHEP {\bf 07} (2002) 051
[\href{http://www.arXiv.org/abs/hep-th/0203163}{{\tt hep-th/0203163}}].

\bibitem{Sarangi:2002yt}
S.~Sarangi and S.~H.~H. Tye,  {\em {Cosmic string production towards the end of
  brane inflation}}, Phys. Lett. {\bf B536} (2002) 185--192
[\href{http://www.arXiv.org/abs/hep-th/0204074}{{\tt hep-th/0204074}}].

\bibitem{Jones:2003da}
N.~T. Jones, H.~Stoica and S.~H.~H. Tye,  {\em {The Production, spectrum and
  evolution of cosmic strings in brane inflation}}, Phys. Lett. {\bf B563}
  (2003) 6--14
[\href{http://www.arXiv.org/abs/hep-th/0303269}{{\tt hep-th/0303269}}].

\bibitem{Copeland:2003bj}
E.~J. Copeland, R.~C. Myers and J.~Polchinski,  {\em {Cosmic F and D strings}},
  JHEP {\bf 06} (2004) 013
[\href{http://www.arXiv.org/abs/hep-th/0312067}{{\tt hep-th/0312067}}].

\bibitem{Schwarz:1995dk}
J.~H. Schwarz,  {\em {An SL(2,Z) multiplet of type IIB superstrings}}, Phys.
  Lett. {\bf B360} (1995) 13--18
  [\href{http://www.arXiv.org/abs/hep-th/9508143}{{\tt hep-th/9508143}}],
[Erratum: Phys. Lett.B364,252(1995)].

\bibitem{Gubser:2004qj}
S.~S. Gubser, C.~P. Herzog and I.~R. Klebanov,  {\em {Symmetry breaking and
  axionic strings in the warped deformed conifold}}, JHEP {\bf 09} (2004) 036
[\href{http://www.arXiv.org/abs/hep-th/0405282}{{\tt hep-th/0405282}}].

\bibitem{Firouzjahi:2006vp}
H.~Firouzjahi, L.~Leblond and S.~H. Henry~Tye,  {\em {The (p,q) string tension
  in a warped deformed conifold}}, JHEP {\bf 05} (2006) 047
[\href{http://www.arXiv.org/abs/hep-th/0603161}{{\tt hep-th/0603161}}].

\bibitem{Polchinski:2004ia}
J.~Polchinski,  {\em {Introduction to cosmic F- and D-strings}}, in {\em
  {String theory: From gauge interactions to cosmology. Proceedings, NATO
  Advanced Study Institute, Cargese, France, June 7-19, 2004}}, pp.~229--253.
\newblock 2004.
\newblock
\href{http://www.arXiv.org/abs/hep-th/0412244}{{\tt hep-th/0412244}}.
\newblock

\bibitem{Chernoff:2014cba}
D.~F. Chernoff and S.~H.~H. Tye,  {\em {Inflation, string theory and cosmic
  strings}}, Int. J. Mod. Phys. {\bf D24} (2015), no.~03, 1530010
[\href{http://www.arXiv.org/abs/1412.0579}{{\tt 1412.0579}}].

\bibitem{Sakellariadou:2004wq}
M.~Sakellariadou,  {\em {A Note on the evolution of cosmic string/superstring
  networks}}, JCAP {\bf 0504} (2005) 003
[\href{http://www.arXiv.org/abs/hep-th/0410234}{{\tt hep-th/0410234}}].

\bibitem{Firouzjahi:2007dp}
H.~Firouzjahi,  {\em {Energy radiation by cosmic superstrings in brane
  inflation}}, Phys. Rev. {\bf D77} (2008) 023532
[\href{http://www.arXiv.org/abs/0710.4609}{{\tt 0710.4609}}].

\bibitem{Csaki:2001jk}
C.~Csaki, N.~Kaloper and J.~Terning,  {\em {Effects of the intergalactic plasma
  on supernova dimming via photon axion oscillations}}, Phys. Lett. {\bf B535}
  (2002) 33--36
[\href{http://www.arXiv.org/abs/hep-ph/0112212}{{\tt hep-ph/0112212}}].

\bibitem{Mirizzi:2009iz}
A.~Mirizzi, J.~Redondo and G.~Sigl,  {\em {Microwave Background Constraints on
  Mixing of Photons with Hidden Photons}}, JCAP {\bf 0903} (2009) 026
[\href{http://www.arXiv.org/abs/0901.0014}{{\tt 0901.0014}}].

\bibitem{Mirizzi:2009nq}
A.~Mirizzi, J.~Redondo and G.~Sigl,  {\em {Constraining resonant photon-axion
  conversions in the Early Universe}}, JCAP {\bf 0908} (2009) 001
[\href{http://www.arXiv.org/abs/0905.4865}{{\tt 0905.4865}}].

\bibitem{Damour:2001bk}
T.~Damour and A.~Vilenkin,  {\em {Gravitational wave bursts from cusps and
  kinks on cosmic strings}}, Phys. Rev. {\bf D64} (2001) 064008
[\href{http://www.arXiv.org/abs/gr-qc/0104026}{{\tt gr-qc/0104026}}].

\bibitem{Damour:2004kw}
T.~Damour and A.~Vilenkin,  {\em {Gravitational radiation from cosmic
  (super)strings: Bursts, stochastic background, and observational windows}},
  Phys. Rev. {\bf D71} (2005) 063510
[\href{http://www.arXiv.org/abs/hep-th/0410222}{{\tt hep-th/0410222}}].

\bibitem{Raffelt:1987im}
G.~Raffelt and L.~Stodolsky,  {\em {Mixing of the Photon with Low Mass
  Particles}}, Phys. Rev. {\bf D37} (1988)
1237.

\bibitem{Yanagida:1987nf}
T.~Yanagida and M.~Yoshimura,  {\em {Resonant Axion - Photon Conversion in the
  Early Universe}}, Phys. Lett. {\bf B202} (1988)
301--306.

\bibitem{Moroi:2018vci}
T.~Moroi, K.~Nakayama and Y.~Tang,  {\em {Axion-photon conversion and effects
  on 21 cm observation}}, Phys. Lett. {\bf B783} (2018) 301--305
[\href{http://www.arXiv.org/abs/1804.10378}{{\tt 1804.10378}}].

\bibitem{Xu:2016vjj}
S.~Xu and B.~Zhang,  {\em {On the origin of the scatter broadening of fast
  radio burst pulses and astrophysical implications}}, Astrophys. J. {\bf 832}
  (2016), no.~2, 199
[\href{http://www.arXiv.org/abs/1608.03930}{{\tt 1608.03930}}].

\bibitem{Martins:1995tg}
C.~J. A.~P. Martins and E.~P.~S. Shellard,  {\em {String evolution with
  friction}}, Phys. Rev. {\bf D53} (1996) 575--579
[\href{http://www.arXiv.org/abs/hep-ph/9507335}{{\tt hep-ph/9507335}}].

\bibitem{Battye:1997hu}
R.~A. Battye, J.~Robinson and A.~Albrecht,  {\em {Structure formation by cosmic
  strings with a cosmological constant}}, Phys. Rev. Lett. {\bf 80} (1998)
  4847--4850
[\href{http://www.arXiv.org/abs/astro-ph/9711336}{{\tt astro-ph/9711336}}].

\bibitem{Pogosian:1999np}
L.~Pogosian and T.~Vachaspati,  {\em {Cosmic microwave background anisotropy
  from wiggly strings}}, Phys. Rev. {\bf D60} (1999) 083504
[\href{http://www.arXiv.org/abs/astro-ph/9903361}{{\tt astro-ph/9903361}}].

\bibitem{Martins:2000cs}
C.~J. A.~P. Martins and E.~P.~S. Shellard,  {\em {Extending the velocity
  dependent one scale string evolution model}}, Phys. Rev. {\bf D65} (2002)
  043514
[\href{http://www.arXiv.org/abs/hep-ph/0003298}{{\tt hep-ph/0003298}}].

\bibitem{Polchinski:2007rg}
J.~Polchinski and J.~V. Rocha,  {\em {Cosmic string structure at the
  gravitational radiation scale}}, Phys. Rev. {\bf D75} (2007) 123503
[\href{http://www.arXiv.org/abs/gr-qc/0702055}{{\tt gr-qc/0702055}}].

\bibitem{Polchinski:2007qc}
J.~Polchinski,  {\em {Cosmic String Loops and Gravitational Radiation}}, in
  {\em {Recent developments in theoretical and experimental general relativity,
  gravitation and relativistic field theories. Proceedings, 11th Marcel
  Grossmann Meeting, MG11, Berlin, Germany, July 23-29, 2006. Pt. A-C}},
  pp.~105--125.
\newblock 2007.
\newblock
\href{http://www.arXiv.org/abs/0707.0888}{{\tt 0707.0888}}.
\newblock

\bibitem{Ringeval:2005kr}
C.~Ringeval, M.~Sakellariadou and F.~Bouchet,  {\em {Cosmological evolution of
  cosmic string loops}}, JCAP {\bf 0702} (2007) 023
[\href{http://www.arXiv.org/abs/astro-ph/0511646}{{\tt astro-ph/0511646}}].

\bibitem{Lorenz:2010sm}
L.~Lorenz, C.~Ringeval and M.~Sakellariadou,  {\em {Cosmic string loop
  distribution on all length scales and at any redshift}}, JCAP {\bf 1010}
  (2010) 003
[\href{http://www.arXiv.org/abs/1006.0931}{{\tt 1006.0931}}].

\bibitem{BlancoPillado:2011dq}
J.~J. Blanco-Pillado, K.~D. Olum and B.~Shlaer,  {\em {Large parallel cosmic
  string simulations: New results on loop production}}, Phys. Rev. {\bf D83}
  (2011) 083514
[\href{http://www.arXiv.org/abs/1101.5173}{{\tt 1101.5173}}].

\bibitem{Blanco-Pillado:2013qja}
J.~J. Blanco-Pillado, K.~D. Olum and B.~Shlaer,  {\em {The number of cosmic
  string loops}}, Phys. Rev. {\bf D89} (2014), no.~2, 023512
[\href{http://www.arXiv.org/abs/1309.6637}{{\tt 1309.6637}}].

\bibitem{Chernoff:2009tp}
D.~F. Chernoff,  {\em {Clustering of Superstring Loops}},
\href{http://www.arXiv.org/abs/0908.4077}{{\tt 0908.4077}}.

\bibitem{Chernoff:2017fll}
D.~F. Chernoff and S.~H.~H. Tye,  {\em {Detection of Low Tension Cosmic
  Superstrings}}, JCAP {\bf 1805} (2018), no.~05, 002
[\href{http://www.arXiv.org/abs/1712.05060}{{\tt 1712.05060}}].

\bibitem{Sousa:2016ggw}
L.~Sousa and P.~P. Avelino,  {\em {Probing Cosmic Superstrings with
  Gravitational Waves}}, Phys. Rev. {\bf D94} (2016), no.~6, 063529
[\href{http://www.arXiv.org/abs/1606.05585}{{\tt 1606.05585}}].

\bibitem{Firouzjahi:2005dh}
H.~Firouzjahi and S.~H.~H. Tye,  {\em {Brane inflation and cosmic string
  tension in superstring theory}}, JCAP {\bf 0503} (2005) 009
[\href{http://www.arXiv.org/abs/hep-th/0501099}{{\tt hep-th/0501099}}].

\bibitem{Kachru:2003sx}
S.~Kachru, R.~Kallosh, A.~D. Linde, J.~M. Maldacena, L.~P. McAllister and S.~P.
  Trivedi,  {\em {Towards inflation in string theory}}, JCAP {\bf 0310} (2003)
  013
[\href{http://www.arXiv.org/abs/hep-th/0308055}{{\tt hep-th/0308055}}].

\end{thebibliography}\endgroup

\end{document}